\documentclass[12pt,draftclsnofoot,onecolumn]{IEEEtran}
\usepackage{amsmath,amsfonts}
\usepackage{algorithmic}
\usepackage{algorithm}
\usepackage{array}
\usepackage[caption=false,font=normalsize,labelfont=sf,textfont=sf]{subfig}
\usepackage{textcomp}
\usepackage{stfloats}
\usepackage{url}
\usepackage{verbatim}
\usepackage{graphicx}
\usepackage{cite}
\usepackage{color}
\usepackage{amssymb}

\begin{document}

\title{Joint Distributed Precoding and Beamforming for RIS-aided Cell-Free Massive MIMO Systems}

\author{Peng Zhang, Jiayi Zhang~\IEEEmembership{Senior Member,~IEEE}, Huahua Xiao, Xiaodan Zhang, \\Derrick Wing Kwan Ng,~\IEEEmembership{Fellow,~IEEE}, and Bo Ai,~\IEEEmembership{Fellow,~IEEE}
\thanks{Copyright (c) 2015 IEEE. Personal use of this material is permitted. However, permission to use this material for any other purposes must be obtained from the IEEE by sending a request to pubs-permissions@ieee.org.}
\thanks{This work was supported in part by the National Key
Research and Development Program of China under Grant 2020YFB1807201,
in part by the National Natural Science Foundation of China under Grant
61971027 and Grant 62221001, in part by the Beijing Natural Science
Foundation under Grant L202013, in part by the Fundamental Research Funds for the Central Universities under Grant 2022JBQY004, in part by the Natural Science
Foundation of Jiangsu Province Major Project under Grant BK20212002, in part by the Key Research Projects of Universities in Guangdong Province under Grant 2021ZDZX1134, and in part by ZTE Industry-University-Institute Cooperation Funds under Grant No. HC-CN-20221202003. D. W. K. Ng is supported by the Australian Research Council's Discovery Projects (DP210102169, DP230100603). \emph{(Corresponding author: Jiayi Zhang and Xiaodan Zhang.)}}
\thanks{P. Zhang, J. Zhang, and B. Ai are with the School of Electronic and Information Engineering, Beijing Jiaotong University, Beijing 100044, P.R. China, and also with the Frontiers Science Center for Smart High-speed Railway System, Beijing Jiaotong University, Beijing 100044, China. (e-mail: jiayizhang@bjtu.edu.cn).}
\thanks{H. Xiao is with ZTE Corporation and  State Key Laboratory of Mobile Network and Mobile Multimedia Technology. (e-mail: xiao.huahua@zte.com.cn).}
\thanks{X. Zhang is with School of Management, Shenzhen Institute of Information Technology, Shenzhen 518172, China (e-mail: zhangxd@sziit.edu.cn).}
\thanks{D. W. K. Ng is with the School of Electrical Engineering and Telecommunications, University of New South Wales, NSW 2052, Australia. (e-mail: w.k.ng@unsw.edu.au).}
}

\maketitle

\begin{abstract}
The amalgamation of cell-free networks and reconfigurable intelligent surface (RIS) has become a prospective technique for future sixth-generation wireless communication systems. In this paper, we focus on the precoding and beamforming design for a downlink RIS-aided cell-free network. The design is formulated as a non-convex optimization problem by jointly optimizing the combining vector, active precoding, and passive RIS beamforming for minimizing the weighted sum of users' mean square error. A novel joint distributed precoding and beamforming framework is proposed to decentralize the alternating optimization method for acquiring a suboptimal solution to the design problem. Finally, numerical results validate the effectiveness of the proposed distributed precoding and beamforming framework, showing its low-complexity and improved scalability compared with the centralized method.
\end{abstract}

\begin{IEEEkeywords}
Reconfigurable intelligent surface, distributed precoding, passive beamforming.
\end{IEEEkeywords}

\section{Introduction}
\IEEEPARstart{T}{h}e skyrocketing demand for improved network capacity, higher user data rates, and seamless connectivity has fueled the evolution of the sixth-generation (6G) wireless communication systems. To satisfy these unparalleled demands, the notion of cell-free massive multiple-input multiple-output (mMIMO) has been proposed. In particular, it introduced a decentralized antenna architecture, featuring an extensive deployment of collaborating access points (APs) to serve multiple users simultaneously \cite{demir2021foundations}. In light of these, various efforts have been devoted. For instance, closed-form capacity lower bounds for the cell-free mMIMO downlink and uplink were derived in \cite{ngo2017cell}. Also, comprehensive analysis of cell-free mMIMO system under different degrees of cooperation among the APs were provided in \cite{bjornson2019making}. Besides, a new framework for scalable cell-free mMIMO systems were proposed in \cite{bjornson2020scalable}.

However, realizing cell-free mMIMO networks in practice presents unique challenges due to their exceedingly high signaling overhead and complexity in network management \cite{masoumi2019performance}. Fortunately, reconfigurable intelligent surface (RIS), an emerging technique, offered a promising solution to enhance network capacity and energy efficiency \cite{hu2021robust}.
As such, the incorporation of RIS as a cost-effective and energy-efficient solution for cell-free mMIMO networks offers tremendous potential for optimizing network capacity and improving the overall performance of wireless communication systems~\cite{shi2022wireless}. To fully leverage the benefits of RISs in cell-free mMIMO networks, the joint design of APs active precoding and RIS passive beamforming is of paramount importance \cite{zhang2021joint,ma2023cooperative,ni2022partially,jin2022ris}. The initial concept of RIS-aided cell-free networks was introduced in \cite{zhang2021joint}. Specifically, the authors in \cite{ma2023cooperative} studied the characteristic of imperfect channel state information (CSI) and solved the weighted sum-rate (WSR) maximization problem in RIS-aided cell-free network. Besides, a partially distributed beamforming design scheme was proposed for RIS-aided cell-free networks in \cite{ni2022partially}. Furthermore, the authors in \cite{jin2022ris} explored the max-min fairness problem, aiming to maximize the minimum achievable rate among all the users in RIS-aided cell-free networks.

In spite of the fruitful results in the literature, existing techniques only address the beamforming design optimization problem at a central processing unit (CPU) in a centralized manner. As the size of network scales up, conventional centralized algorithms are unable to cope with the need for timely and scalable signal processing. Furthermore, due to the decentralized characteristics of APs and users, acquiring the fully-known CSI becomes almost impossible for a CPU. On the other hand, the conventional distributed algorithms are only applicable to multi-cell or broadcast wireless communication systems \cite{choi2012distributed,bjornson2010cooperative}, while the precoding design of RIS-assisted cell-free systems involves the joint optimization of active precoding vectors and phase shift matrix of RISs.
This research gap calls for further exploration.


To fully unleash the potential gains brought by the cell-free mMIMO distributed architecture, we propose a joint distributed precoding and beamforming framework for the RIS-aided cell-free network to reduce the computational complexity and improve the scalability of the network. The main contributions of this paper are summarized below:
\begin{itemize}
	\item We study the precoding and beamforming design for a downlink RIS-aided cell-free network. The design is formulated as a weighted sum of users' MSE minimization problem that jointly optimizes the combining vector, active precoding, and passive RIS beamforming design.
	\item We propose a distributed precoding and beamforming framework by decentralizing the alternating optimization problem to each AP with a significantly lower computational complexity when compared to centralized algorithms.
	\item Numerical results verify that the performance of proposed distributed precoding and beamforming framework closely approaches that of the centralized method. Important insights related to the impacts of key system parameters (i.e., the transmit power of APs and the number of RIS elements) are also revealed.
\end{itemize}


\section{System Model}
\begin{figure}[!t]
	\centering
	\includegraphics[width=0.7\textwidth]{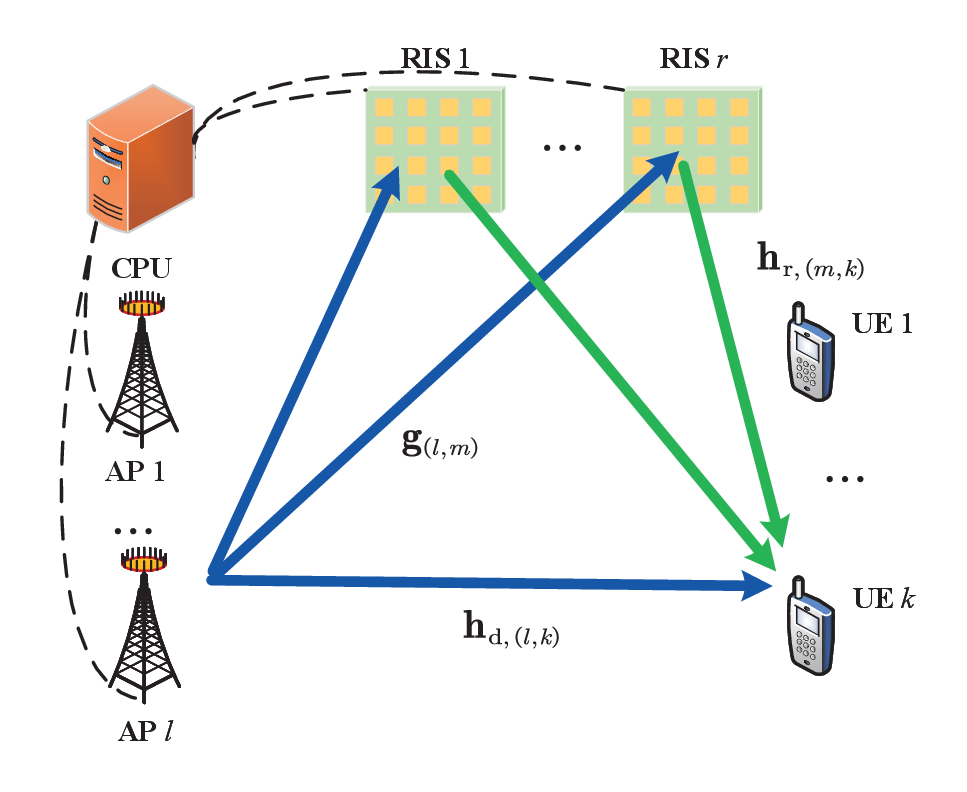}
	\caption{System block diagram of a general RIS-aided cell-free network.}
	\label{fig:system_model}
\end{figure}
Herein, we investigate an RIS-aided cell-free network as Fig.~\ref{fig:system_model}, comprising $L$ APs, $R$ RISs, and $K$ multi-antenna users \cite{demir2021foundations}. The number of antennas at the $l$-th AP, $l\in\mathcal{L} =\left\{ 1,2,\cdots ,L \right\} $, and that at the $k$-th user, $k\in\mathcal{K} =\left\{ 1,2,\cdots ,K \right\} $ are $N_{t}$ and $N_{r}$, respectively. The number of elements at the $r$-th RIS, $r\in\mathcal{R} =\left\{ 1,2,\cdots ,R \right\} $ is $M$.

\subsection{Downlink Transmission}
The equivalent channel of the $l$-th AP to the user $k$ can be written as \cite{zhang2021joint}
\begin{equation}
	\label{equ_channel}
	\mathbf{H}_{\left( l,k \right)}^{\mathsf{H}}=\mathbf{H}_{\mathrm{d},\left( l,k \right)}^{\mathsf{H}}+\sum_{r\in \mathcal{R}}{\mathbf{H}_{\mathrm{r},\left( r,k \right)}^{\mathsf{H}}\mathbf{\Theta }_r\mathbf{G}_{\left( l,r \right)}},
\end{equation}
where $\mathbf{H}_{\mathrm{d},\left( l,k \right)}\in \mathbb{C} ^{N_t\times N_r}$, $\mathbf{H}_{\mathrm{r},\left( r,k \right)}\in \mathbb{C} ^{M\times N_r}$, and $\mathbf{G}_{\left( l,r \right)}\in \mathbb{C} ^{M\times N_t}$ represent the direct downlink channel from AP $l$-to-user $k$, from RIS $r$-to-user $k$, and from AP $l$-to-RIS $r$, respectively; $\mathbf{\Theta }_r=\mathrm{diag}\left\{ \phi _{r,1},\cdots ,\phi _{r,M} \right\}$ represents the phase shift matrix at RIS $r$, where $\phi _{r,m}$ denotes the phase shift of $m$-th element of the $r$-th RIS, and $\left| \phi _{r,m} \right|\le 1, \forall r\in \mathcal{R} , \forall m\in \left\{ 1,2,\cdots ,M \right\}$  for an ideal RIS case \cite{zhang2021joint}. Assuming that the local CSI\footnote{We will consider the impact of imperfect CSI in our future works.} of the $l$-th AP, i.e.  $\mathbf{H}_{\left( l,k \right)}, \forall k \in \mathcal{K}$, is perfectly known at the $l$-th AP \cite{zhang2021joint}.

To simplify the equivalent channel $\mathbf{H}_{\left( l,k \right)}^{\mathsf{H}}$, we define $\mathbf{h}_k=[ \mathbf{H}_{\mathrm{r},\left( 1,k \right)}^{\mathsf{T}},\dots ,\mathbf{H}_{\mathrm{r},\left( R,k \right)}^{\mathsf{T}} ] ^{\mathsf{T}}\in \mathbb{C} ^{MR\times 1} $, $\mathbf{\Theta }=\mathrm{diag}\left\{ \mathbf{\Theta }_1,\dots ,\mathbf{\Theta }_R \right\} \in \mathbb{C} ^{MR\times MR}
$, and $\mathbf{G}_l=[ \mathbf{G}_{\left( l,1 \right)}^{\mathsf{T}},\dots ,\mathbf{G}_{\left( l,R \right)}^{\mathsf{T}} ] ^{\mathsf{T}}\in \mathbb{C} ^{MR\times N_t}$. Therefore, the equivalent channel $\mathbf{H}_{\left( l,k \right)}^{\mathsf{H}}$ in \eqref{equ_channel} can be expressed as
\begin{equation}
	\label{equ:H_lk}
	\mathbf{H}_{\left( l,k \right)}^{\mathsf{H}}=\mathbf{H}_{\mathrm{d},\left( l,k \right)}^{\mathsf{H}}+\mathbf{h}_{k}^{\mathsf{H}}\mathbf{\Theta G}_l.
\end{equation}

Let $s_k \in\mathbb{C}$ denote the transmitted symbols to user $k$, where $\mathbb{E} \{ \left| s_k \right|^2 \} =1$, $\forall k\in\mathcal{K}$. In the downlink, the transmission symbol $s_k$ is initially precoded by $\mathbf{f}_{\left( l,k \right)}\in \mathbb{C} ^{N_t\times 1}$ at the $l$-th AP and the precoded symbol $\mathbf{x}_l\in\mathbb{C}^{N_t\times 1}$ at the $l$-th AP can be expressed as
\begin{equation}
	\mathbf{x}_l=\sum_{k\in \mathcal{K}}{\mathbf{f}_{\left( l,k \right)}s_k},
\end{equation}
where
\begin{equation}
	\sum_{k\in \mathcal{K}}{\left\| \mathbf{f}_{\left( l,k \right)} \right\|}^2\le P_{l,\max}, \forall l\in \mathcal{L},
\end{equation}
where $P_{l,\max}$ represents the maximum transmit power of AP~$l$. The received signal from user $k$ can be expressed as
\begin{equation}
	\label{equ:y_k}
	\begin{split}
		\mathbf{y}_k&=\sum_{i\in \mathcal{K}}{\sum_{l\in \mathcal{L}}{\mathbf{H}_{\left( l,k \right)}^{\mathsf{H}}\mathbf{f}_{\left( l,i \right)}s_i}}+\mathbf{z}_k
		\\
		&\overset{\left( a \right)}{=}\sum_{i\in \mathcal{K}}{\mathbf{H}_{k}^{\mathsf{H}}\mathbf{f}_is_i}+\mathbf{z}_k
		\\
		&=\underset{\text{Desired signal}}{\underbrace{\mathbf{H}_{k}^{\mathsf{H}}\mathbf{f}_ks_k}}+\underset{\text{Inter-user interference}}{\underbrace{\sum_{i\in \mathcal{K} \backslash k}{\mathbf{H}_{k}^{\mathsf{H}}\mathbf{f}_is_i}}}+\underset{\text{Noise}}{\underbrace{\mathbf{z}_k}},
	\end{split}
\end{equation}
where $\left( a\right) $ holds by defining $\mathbf{H}_k=[ \mathbf{H}_{\left( 1,k \right)}^{\mathsf{T}},\dots ,\mathbf{H}_{\left( L,k \right)}^{\mathsf{T}} ] ^{\mathsf{T}}\in \mathbb{C} ^{LN_t\times N_r}$ and $\mathbf{f}_k=[ \mathbf{f}_{\left( 1,k \right)}^{\mathsf{T}},\dots ,\mathbf{f}_{\left( L,k \right)}^{\mathsf{T}} ] ^{\mathsf{T}}\in \mathbb{C} ^{LN_t\times 1}$, and $\mathbf{z}_k\in\mathbb{C}^{N_r\times 1}$ denotes the additive white Gaussian noise (AWGN) at user $k$ with elements distributed as $\mathcal{CN}\left(\mathbf{0}, \sigma_k^2\mathbf{I}_{N_r} \right) $.

After receiving $\mathbf{y}_k$ as in \eqref{equ:y_k}, user $k$ adopts a combining vector $\mathbf{u}_k\in\mathbb{C}^{N_r\times 1}$ to combine $\mathbf{y}_k$. The signal-to-interference-plus-noise ratio (SINR) is represented by
\begin{equation}
	\label{equ:SINR}
	\mathsf{SINR}_k=\frac{\left| \mathbf{u}_{k}^{\mathsf{H}}\mathbf{H}_{k}^{\mathsf{H}}\mathbf{f}_k \right|^2}{\sum_{i\in \mathcal{K} \backslash k}{\left| \mathbf{u}_{k}^{\mathsf{H}}\mathbf{H}_{k}^{\mathsf{H}}\mathbf{f}_i \right|}^2+\left\| \mathbf{u}_{k}^{\mathsf{H}} \right\| ^2\sigma _{k}^{2}}, \forall k\in\mathcal{K}.
\end{equation}
Due to the non-convex expression of SINR above, the joint design of precoding vectors $\mathbf{f}_{\left( l,k \right)}$, the combining vectors $\mathbf{u}_k $ and the phase shift matrixes of RIS $\mathbf{\Theta}_r$ is generally intractable. Fortunately, inspired by transceiver design algorithm and well-know relation between the $k$-th user's mean square error (MSE) $\mathsf{MSE}_k$ and the rate $R_k$, expressed as, $R_k=\log\left( \mathsf{MSE}_k^{-1}\right) $ in \cite{shi2011iteratively}, we can address the problem of maximizing WSR by minimizing the weighted sum of users' MSE, as described below.

\subsection{Problem Fomulation}
The mean square error (MSE) at user $k$ is expressed as
\begin{equation}
	\label{equ:MSE}
	\begin{split}
		\mathsf{MSE}_k&=\mathbb{E} \left\{ \left| \mathbf{u}_{k}^{\mathsf{H}}\mathbf{y}_k-s_k \right|^2 \right\}
		\\
		&=\mathbb{E} \left\{ \mathbf{u}_{k}^{\mathsf{H}}\mathbf{y}_k\mathbf{y}_{k}^{\mathsf{H}}\mathbf{u}_k \right\} -\mathbb{E} \left\{ 2\mathfrak{R} \mathfrak{e} \left\{ \mathbf{u}_{k}^{\mathsf{H}}\mathbf{y}_ks_{k}^{\mathsf{H}} \right\} \right\} +1
		\\
		&=\sum_{i\in \mathcal{K}}{\left| \mathbf{u}_{k}^{\mathsf{H}}\mathbf{H}_{k}^{\mathsf{H}}\mathbf{f}_i \right|^2}-2\mathfrak{R} \mathfrak{e} \left\{ \mathbf{u}_{k}^{\mathsf{H}}\mathbf{H}_{k}^{\mathsf{H}}\mathbf{f}_k \right\} +\left\| \mathbf{u}_k \right\| ^2\sigma _{k}^{2}+1.
	\end{split}
\end{equation}
Therefore, the minimization of the weighted sum of users' MSE problem can be originally formulated as
\begin{subequations}
	\label{equ:P1}
\begin{align}
	\left( \mathrm{P}1 \right) \,\,&\min_{\mathbf{u}_k,\mathbf{F},\mathbf{\Theta }} \,\,\sum_{k\in \mathcal{K}}{\omega _k\mathsf{MSE}_k}
	\\
	\mathrm{s}.\mathrm{t}.   &\sum_{k\in \mathcal{K}}{\left\| \mathbf{f}_{\left( l,k \right)} \right\| ^2}\le P_{l,\max}, \forall l\in \mathcal{L},
	\\
	&\left| \phi _{r,m} \right|\le 1, \forall r\in \mathcal{R} , \forall m\in \left\{ 1,2,\cdots ,M \right\}.
\end{align}
\end{subequations}
where $\mathbf{F}\triangleq \left[ \mathbf{f}_1,\dots ,\mathbf{f}_K \right] \in \mathbb{C} ^{LN_t\times K}$ denotes the precoding matrix of APs and $\mathbf{\Theta }=\mathrm{diag}\left\{ \mathbf{\Theta }_1,\dots ,\mathbf{\Theta }_R \right\}$ represents RIS-based beamforming achieved by determining the phase shifts of all the elements of RISs.

Note that the weighted sum of users' MSE above is convex with respect to the transmit and the receive schemes separately, but not jointly convex that hinders the joint optimization of $\left\{ \mathbf{u}_k \right\} _{\forall k\in \mathcal{K}},\mathbf{F},\mathbf{\Theta }$ for obtaining the globally optimal solution. Hence, as a compromise, we aim to acquire a local optimum of the sum weighted MSE minimization problem by exploiting alternating optimization.

\section{Proposed Joint Distributed Framework}

\subsection{Combining: Fix $(\mathbf{F},\mathbf{\Theta})$ and Solve $\mathbf{u}_k$}
For a given $(\mathbf{F},\mathbf{\Theta})$ and omitted the unrelated terms, the equivalent MSE minimum problem (P1) in \eqref{equ:P1} can be reformulated as
\begin{equation}
	\label{equ:u_k}
	\min_{\mathbf{u}_k} \,\,\mathsf{MSE}_k=\mathbf{u}_{k}^{\mathsf{H}}\left( \mathbf{W}_k+\sigma _{k}^{2}\mathbf{I}_{N_r} \right) \mathbf{u}_k-2\mathfrak{R} \mathfrak{e} \left\{ \mathbf{u}_{k}^{\mathsf{H}}\mathbf{a}_k \right\} ,\forall k\in \mathcal{K},
\end{equation}
where $\mathbf{a}_k\triangleq \mathbf{H}_{k}^{\mathsf{H}}\mathbf{f}_k$ and $\mathbf{W}_k$ is defined as
\begin{equation}
	\label{W_k}
	\mathbf{W}_k\triangleq \sum_{i\in \mathcal{K}}{\left( \mathbf{H}_{k}^{\mathsf{H}}\mathbf{f}_i \right) \left( \mathbf{H}_{k}^{\mathsf{H}}\mathbf{f}_i \right) ^{\mathsf{H}}}.
\end{equation}
The combining vector $\mathbf{u}_k$ that minimizes \eqref{equ:u_k} corresponds to the well-known MMSE (minimum MSE) receiver and can be expressed as
\begin{equation}
	\label{equ:u_k^opt}
	\mathbf{u}_k=\left( \mathbf{W}_k+\sigma _{k}^{2}\mathbf{I}_{N_r} \right) ^{-1}\mathbf{a}_k.
\end{equation}
Note that user $k$ can compute $\mathbf{u}_k$ locally as shown in \eqref{equ:u_k^opt}, if the knowledge of $\mathbf{W}_k$ and the effective channel $\mathbf{a}_k$ is available.
\subsection{Active Precoding: Fix $(\mathbf{u}_k, \mathbf{\Theta})$ and Solve $\mathbf{f}_{\left( l,k \right)}$}
For a fixed pair of $(\mathbf{u}_k, \mathbf{\Theta})$ and omitted the irrelevant terms, the equivalent MSE minimum problem (P1) in \eqref{equ:P1} can be reformulated as
\begin{subequations}
	\label{equ:p2}
	\begin{align}
		\left( \mathrm{P}2 \right) \,\,&\min_{\{ \mathbf{f}_{\left( l,k \right)} \} _{\forall k\in \mathcal{K}}} \,\,g_1\left( \mathbf{F} \right)
		\\
		\mathrm{s}.\mathrm{t}.   &\sum_{k\in \mathcal{K}}{\left\| \mathbf{f}_{\left( l,k \right)} \right\| ^2}\le P_{l,\max}, \forall l\in \mathcal{L},
	\end{align}
\end{subequations}
where
\begin{equation}
	\begin{split}
		\label{equ:g_1_org}
		g_1\left( \mathbf{F} \right) &=\sum_{k\in \mathcal{K}}{\sum_{i\in \mathcal{K}}{\omega _k\left| \sum_{l\in \mathcal{L}}{\mathbf{u}_{k}^{\mathsf{H}}\mathbf{H}_{\left( l,k \right)}^{\mathsf{H}}\mathbf{f}_{\left( l,i \right)}} \right|^2}}
		\\
		&-2\mathfrak{R} \mathfrak{e} \left\{ \sum_{k\in \mathcal{K}}{\sum_{l\in \mathcal{L}}{\omega _k\mathbf{u}_{k}^{\mathsf{H}}\mathbf{H}_{\left( l,k \right)}^{\mathsf{H}}\mathbf{f}_{\left( l,k \right)}}} \right\}.
	\end{split}
\end{equation}
After simplifying the presentation, $g_1\left( \mathbf{F} \right)$ as in \eqref{equ:g_1_org} can be expressed as
\begin{equation}
	g_1\left( \mathbf{F} \right) =\mathrm{tr}\left( \mathbf{F}^{\mathsf{H}}\mathbf{\Lambda F} \right) -2\mathfrak{R} \mathfrak{e} \left\{ \mathrm{tr}\left( \mathbf{\Omega B}^{\mathsf{H}}\mathbf{F} \right) \right\}.
\end{equation}
where $\mathbf{\Omega }\triangleq \mathrm{diag}\left\{ \omega _1,\dots ,\omega _K \right\} \in \mathbb{R} ^{K\times K}$, $\mathbf{b}_{\left( l,k \right)}\triangleq\mathbf{H}_{\left( l,k \right)}\mathbf{u}_k\in \mathbb{C} ^{N_t\times 1}$, $\mathbf{b}_k\triangleq [ \mathbf{b}_{\left( 1,k \right)}^{\mathsf{T}},\dots ,\mathbf{b}_{\left( L,k \right)}^{\mathsf{T}} ] ^{\mathsf{T}}\in \mathbb{C} ^{LN_t\times 1}$, $\mathbf{B}\triangleq \left[ \mathbf{b}_1,\dots ,\mathbf{b}_K \right] \in \mathbb{C} ^{LN_t\times K}$, and $\mathbf{\Lambda }\triangleq \sum_{k\in \mathcal{K}}{\omega _k\mathbf{b}_k\mathbf{b}_{k}^{\mathsf{H}}}\in \mathbb{C} ^{LN_t\times LN_t}$.

Therefore, for each AP $l$ and for each user $k$, the first-order optimality condition of \eqref{equ:p2} can be denoted as
\begin{equation}
	\nabla _{\mathbf{f}_{\left( l,k \right)}}\left( g_1\left( \mathbf{F} \right) -\sum_{\ell \in \mathcal{L}}{\lambda _{\ell}\left( \sum_{i\in \mathcal{K}}{\left\| \mathbf{f}_{\left( \ell ,i \right)} \right\| ^2}-P_{\ell ,\max} \right)} \right) =\mathbf{0},
\end{equation}
where the dual variables $\{\lambda_{\ell}\geq0\}_{\forall \ell\in\mathcal{L}}$ are introduced as the Lagrange multipliers associated with the per-AP power constraints and can be optimized via the bisection method. Finally, the distributed active precoding can be expressed as
\begin{equation}
	\label{equ:f_lk}
	\mathbf{f}_{\left( l,k \right)}=\left( \left[ \mathbf{\Lambda } \right] _{ll}+\lambda _l\mathbf{I}_{N_t} \right) ^{-1}\left( \omega _k\mathbf{b}_{\left( l,k \right)}-\sum_{\ell \in \mathcal{L} \backslash l}{\left[ \mathbf{\Lambda } \right] _{l\ell}\mathbf{f}_{\left( \ell ,k \right)}} \right).
\end{equation}

Observe that the item $\sum_{\ell \in \mathcal{L} \backslash l}{\left[ \mathbf{\Lambda } \right] _{l\ell}\mathbf{f}_{\left( \ell ,k \right)}}$ in \eqref{equ:f_lk} involves information regarding the precoding vectors employed by other APs for user $k$. Knowledge of such inter-AP interactions at each AP $l$ is required for the iterative adjustment of the distributed precoding solution. Consequently, excluding $\sum_{\ell \in \mathcal{L} \backslash l}{\left[ \mathbf{\Lambda } \right] _{l\ell}\mathbf{f}_{\left( \ell ,k \right)}}$ from \eqref{equ:f_lk} results in the suboptimal {\it local MMSE (L-MMSE)} precoding vector \cite{atzeni2020distributed}.

\begin{equation}
	\label{equ:lmmse}
	\mathbf{f}_{\left( l,k \right)}=\omega _k\left( \left[ \mathbf{\Lambda } \right] _{ll}+\lambda _l\mathbf{I}_{N_t} \right) ^{-1}\mathbf{b}_{\left( l,k \right)}.
\end{equation}
Note that L-MMSE precoding in \eqref{equ:lmmse} has the same form as the L-MMSE precoding in conventional cell-free mMIMO system, as shown in \cite[(6.25)]{demir2021foundations}. It only requires knowledge of the local CSI of the $l$-th AP, i.e.,  $\mathbf{H}_{\left( l,k \right)}, \forall k \in \mathcal{K}$, the channel and has no additional CSI exchange required which is locally optimal~\cite{demir2021foundations}.
\subsection{Passive Beamforming: Fix $(\mathbf{u}_k, \mathbf{F})$ and Solve $\mathbf{\Theta}$}
To simplify the subproblem, we introduce the combined downlink equivalent channel $\mathbf{u}_{k}^{\mathsf{H}}\mathbf{H}_{\left( l,k \right)}^{\mathsf{H}}$ as
\begin{align}
		&\mathbf{u}_{k}^{\mathsf{H}}\mathbf{H}_{\left( l,k \right)}^{\mathsf{H}}=\mathbf{u}_{k}^{\mathsf{H}}\mathbf{h}_{\mathrm{d},\left( l,k \right)}^{\mathsf{H}}+\mathbf{u}_{k}^{\mathsf{H}}\mathbf{h}_{k}^{\mathsf{H}}\mathbf{\Theta g}_l\notag
		\\
		&\overset{\left( b \right)}{=}\mathbf{c}_{\mathrm{d},\left( l,k \right)}^{\mathsf{H}}+\mathbf{c}_{k}^{\mathsf{H}}\mathbf{\Theta g}_l
		\overset{\left( c \right)}{=}\mathbf{c}_{\mathrm{d},\left( l,k \right)}^{\mathsf{H}}+\mathbf{\Phi }\mathrm{diag}\left\{ \mathbf{c}_{k}^{\mathsf{H}} \right\} \mathbf{g}_l,
\end{align}
where $\left( b\right) $ holds by defining $\mathbf{c}_{\mathrm{d},\left( l,k \right)}\triangleq \mathbf{h}_{\mathrm{d},\left( l,k \right)}\mathbf{u}_k\in \mathbb{C} ^{N_t\times 1}$ and $\mathbf{c}_k=\mathbf{h}_k\mathbf{u}_k\in \mathbb{C} ^{MR\times 1}$, and $\left( c\right) $ holds by defining $\mathbf{\Phi }\triangleq \mathbf{1}_{RM}^{\mathsf{T}}\mathbf{\Theta }\in \mathbb{C} ^{1\times RM}$.

Based on the given $(\mathbf{u}_k, \mathbf{F})$ and omitted the unrelated terms, the passive beamforming design problem at the RISs can be expressed as
\begin{subequations}
	\label{equ:p3}
	\begin{align}
		\left( \mathrm{P}3 \right) \,\,&\min_{\mathbf{\Phi }} \,\,g_2\left( \mathbf{\Phi } \right)
		\\
		\label{equ:convex_c}\mathrm{s}.\mathrm{t}. &\left| \phi _{r,m} \right|\le 1, \forall r\in \mathcal{R} , \forall m\in \left\{ 1,2,\cdots ,M \right\}
	\end{align}
\end{subequations}
where
\begin{align}
	\label{equ:g_2}
	g_2\left( \mathbf{\Phi } \right) &=\mathbf{\Phi \Sigma \Phi }^{\mathsf{H}}+2\mathfrak{R} \mathfrak{e} \left\{ \mathbf{\Phi U} \right\}, \\
	\mathbf{\Sigma }&\triangleq\sum_{k\in \mathcal{K}}{\sum_{i\in \mathcal{K}}{\omega _k\left( \sum_{l\in \mathcal{L}}{\mathbf{d}_{\left( l,k \right)}\mathbf{f}_{\left( l,i \right)}} \right) \left( \sum_{l\in \mathcal{L}}{\mathbf{d}_{\left( l,k \right)}\mathbf{f}_{\left( l,i \right)}} \right) ^{\mathsf{H}}}},
\end{align}
and
\begin{equation}
	\begin{split}
		\mathbf{U}&\triangleq \sum_{k\in \mathcal{K}}{\sum_{i\in \mathcal{K}}{\omega _k\left( \sum_{l\in \mathcal{L}}{\mathbf{d}_{\left( l,k \right)}\mathbf{f}_{\left( l,i \right)}} \right)\! \left( \sum_{l\in \mathcal{L}}{\mathbf{c}_{\mathrm{d},\left( l,k \right)}^{\mathsf{H}}\mathbf{f}_{\left( l,i \right)}} \right) ^{\mathsf{H}}}}
		\\
		&-\sum_{k\in \mathcal{K}}{\sum_{l\in \mathcal{L}}{\omega _k\mathbf{d}_{\left( l,k \right)}\mathbf{f}_{\left( l,k \right)}}},
	\end{split}
\end{equation}
where $\mathbf{d}_{\left( l,k \right)}\triangleq \mathrm{diag}\left\{ \mathbf{c}_{k}^{\mathsf{H}} \right\} \mathbf{g}_l\in \mathbb{C} ^{MR\times N_t}$. The subproblem in (P3) can be solved utilizing alternating direction method of multipliers (ADMM) \cite{boyd2011distributed}. However, employing ADMM in (P3) problem involves computationally intensive matrix inversion for $\mathbf{U}$ with a complexity order of $\mathcal{O}\left( R^3M^3\right) $ \cite{zhang2021joint}. Since the RIS element number $M$ is usually large in practice, the complexity of adopting ADMM is exceedingly high. To facilitate its implementation, a low-complexity method based on the {\it primal-dual-subgradient (PDS)} can be exploited to obtain the solution \cite{zhang2021joint}, which is omitted here for brevity.

\subsection{Algorithm Implementation}
\begin{algorithm}[t]
	\caption{Joint Distributed Precoding and Beamforming Framework}
	\label{opt_alg}
	\begin{algorithmic}
			\REQUIRE $R$, $K$, $M$, $L$, $N_t$, $N_r$
			\ENSURE Combining vector $\{\mathbf{u}_k\}_{\forall k\in\mathcal{K}}$, active precoding $\{ \mathbf{f}_{\left( l,k \right)} \} _{\forall k\in \mathcal{K}}$, and passive beamforming~$\mathbf{\Theta}$
			\STATE The APs acquire CSIs and then feed back to the CPU.
			\WHILE {not converge}
			\STATE AP $l, \forall l\in\mathcal{L}$, initializes/receives $\mathbf{u}_k^{(i-1)}$, $\mathbf{\Phi}^{(i-1)}$, ${\left[ \mathbf{\Lambda } \right] _{l\ell}\mathbf{f}_{\left( \ell ,k \right)}} $, $\forall \ell \in\mathcal{L}\backslash l$;
			\STATE AP $l, \forall l\in\mathcal{L}$, updates $\{ \mathbf{f}_{\left( l,k \right)} \} _{\forall k\in \mathcal{K}}$ with \eqref{equ:f_lk};
			\STATE The CPU receives $\{ \mathbf{f}_{\left( l,k \right)} \} _{\forall l\in\mathcal{L}, k\in \mathcal{K}}$;
			\STATE The CPU updates $\mathbf{u}_k^{(i)}$ with \eqref{equ:u_k^opt};
			\STATE The CPU updates $\mathbf{\Theta}^{(i)}$ with solving problem (P3);
			\STATE $i\gets i+1;$
			\ENDWHILE
	\end{algorithmic}
\end{algorithm}
The proposed joint distributed precoding and beamforming framework is summarized in Algorithm \ref{opt_alg}. First, each AP $l$ first obtains the channel matrices $\mathbf{H}_{\left(l,k \right)}$ and forwards them to the CPU via backhaul signaling. Then, each AP locally computes its active precoding $\{ \mathbf{f}_{\left( l,k \right)} \} _{\forall k\in \mathcal{K}}$ with \eqref{equ:f_lk} in a decentralized manner and forwards them to the CPU via dedicated out-of-band backhaul links, while active precoding $\{ \mathbf{f}_{\left( l,k \right)} \} _{\forall k\in \mathcal{K}}$ for each AP is computed at the CPU for conventional centralized precoding design \cite{zhang2021joint}. In particular, the CPU computes the combining vectors $\{\mathbf{u}_k\}_{\forall k\in\mathcal{K}}$ as in \eqref{equ:u_k^opt} and the passive beamforming $\mathbf{\Theta}$ as in \eqref{equ:p3}. Subsequently, AP $l$ locally obtain convergent precoding vectors $\{ \mathbf{f}_{\left( l,k \right)} \} _{\forall k\in \mathcal{K}}$ by itself and the CPU feeds back RIS-specific passive beamforming matrices $\{\mathbf{\Theta}_r\}_{\forall r\in\mathcal{R}}$ to each RIS. Lastly, each user $k$ acquires its combining vector $\{\mathbf{u}_k\}_{\forall k\in\mathcal{K}}$ as in \eqref{equ:u_k^opt}. Then, a simple signaling overhead analysis is provide as follows.

The signaling overhead of backhaul signaling requires conveying $N_rN_tLK$ symbols. Besides, the required signal for updating at the APs and CPU are $KN_r+RM+(L-1)KN_t$ and $LKN_t$ symbols, respectively. Therefore, the total signaling overhead of the proposed framework after $I_{\rm o}$ iterations is $N_rN_tLK+I_{\rm o}(KN_r+RM+(2L-1)KN_t)$ symbols. In contrast, the total signaling overhead of the conventional centralized one in \cite{zhang2021joint} after $I_{\rm o}$ iterations is $N_rN_tLK+I_{\rm o}(KN_r+RM+2LKN_t)$ symbols. Compared with the conventional centralized algorithm, the proposed framework reduces the signaling overhead by $I_{\rm o}KN_t$ symbols.

\subsection{Computational Complexity}
The overall computational complexities of the proposed framework are mainly comprised of the updates of the variables $\{ \mathbf{u}_{k} \} _{\forall k\in \mathcal{K}}$, $\{ \mathbf{f}_{\left( l,k \right)} \} _{\forall k\in \mathcal{K}}$, and $\{\mathbf{\Theta}_r\}_{\forall r\in\mathcal{R}}$. The computational complexity of the distributed precoding design after $I_{\rm o}$ iterations is $\mathcal{O}\left(I_{\rm o}LN_t^3 \right) $, while that the computational complexity of solving \eqref{equ:u_k^opt} is $\mathcal{O}\left( K N_t^3\right) $. On the other hand, if the PDS method is adopted, the computational complexity of solving \eqref{equ:p3} after $I_{\rm p}$ iterations is $\mathcal{O}\left( I_{\rm p} \left(R^2 M^2 + RM\right)\right)  $. Therefore, the overall computational complexity of the proposed joint distributed precoding and beamforming framework after $I_{\rm a}$ iterations is $\mathcal{O}\left(I_{\rm a}\left( K N_t^3 + I_{\rm o}LN_t^3 +I_{\rm p} \left(R^2 M^2 + RM\right) \right)  \right) $. In contrast, the computational complexity of centralized active precoding strategies after $I_{\rm o}$ iterations  is $\mathcal{O}\left(I_{\rm o}L^3 N_t^3 \right) $, where the term $L^3N_t^3$ follows from the $\left(LN_t \times LN_t \right) $-dimensional matrix inversion. The distributed precoding substantially reduces the computational complexity and improves scalability, since the total number of AP antennas in the network $LN_t$ is usually huge in RIS-assisted cell-free massive MIMO deployment.
\section{Numerical Results and Discussion}
For simplicity, we consider a three-dimensional (3D) scenario with $L=5$, $K=4$, $N_t=3$, $N_r=2$, $M=100$, $P_{l,\max}=0{\text{ dBm}}$, $\sigma_k^2 = -80{\text{ dBm}}, \forall k\in\mathcal{K}$, where the two RISs are separately mounted on the two distant building facades, which are tall enough to establish extra reflection links. The $l$-th AP and the two RISs are located at $\left(40(l-1) {\text{ m}}, -50{\text{ m}}, 3{\text{ m}} \right) $, $\left(60{\text{ m}}, 10{\text{ m}}, 6{\text{ m}} \right) $, and $\left(100{\text{ m}}, 10{\text{ m}}, 6{\text{ m}} \right) $, respectively \cite{zhang2021joint}. Moreover, we consider the same settings of both the large-scale fading model and the small-scale fading model as those in \cite{zhang2021joint}. Besides, $\{\mathbf{u}_k\}_{\forall k\in\mathcal{K}}$ is initialized by setting all of its elements to one,  $\mathbf{F}$ is initialized with identical power and random phases, and $\mathbf{\Theta}$ is initialized by random values satisfying the constraint in \eqref{equ:P1} in the proposed algorithm.

In the following figures, the ``No RIS" curve represents the conventional cell-free network without RIS implementation, while the ``random phase shift" curve is defined as a scenario where all the phase shifts of RIS elements are randomly set.

\subsection{Convergence}
\begin{figure}[!t]
	\centering
	\includegraphics[width=0.7\textwidth]{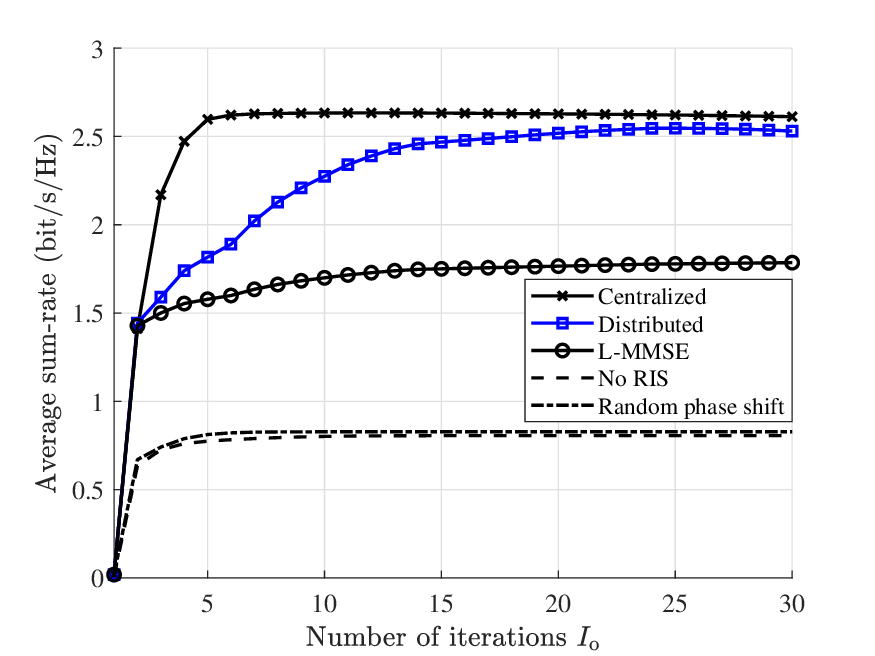}
	\caption{Average sum-rate versus the number of iteration.}
	\label{fig:WSRvsIter}
\end{figure}
To illustrate the convergence of the proposed algorithm, we depict the average sum-rate (ASR) versus the number of iterations $I_{\rm o}$ in Fig. \ref{fig:WSRvsIter}. The results demonstrate that the proposed framework, the centralized case, and the L-MMSE case can converge within 20 iterations, 5 iterations, and 15 iterations, respectively, on average. Since the conventional cell-free network without RIS and the scheme ``Random phase shift" do not need to address the RIS precoding, which can converge within 5 iterations. It can be observed that the performance of the proposed scheme approaches that of the centralized one, thanks to the designed optimization framework. However, compared with the centralized method, the proposed algorithm requires more iterations.

\subsection{The Impact of Key System Parameters}
\subsubsection{Transmit Power of the APs}
\begin{figure}[!t]
	\centering
	\includegraphics[width=0.7\textwidth]{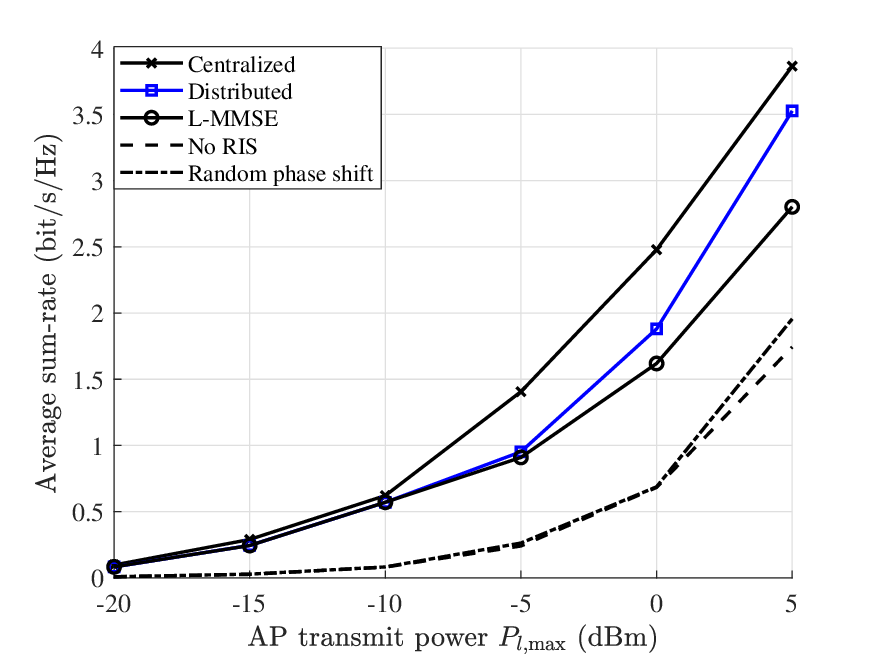}
	\caption{Average sum-rate versus the maximum transmit power $P_{l,\max}$ with $I_{\rm o} = 20$.}
	\label{fig:WSRvsP}
\end{figure}

The ASR versus the AP transmit power with $I_{\rm o} = 20$ is depicted in Fig. \ref{fig:WSRvsP}. We can observe that with the increases of the AP transmit power, the ASR improve rapidly in all the cases. In particular, the proposed distributed scheme scales with the transmit power similarly as the centralized one and approaching the performance of the latter due to the designed optimization framework. Besides, the performance of the ``L-MMSE" is lower that the ``Distributed" one, this is because the L-MMSE precoding employs only local information and does not exchange any information between the APs. Therefore, the WSR is lower that the proposed case. Moreover, when the AP's transmit power is insufficient, the reflected signals by the RISs are weak such that RISs barely have any contribution to the performance improvement. Indeed, the performance gain provided by RISs is significant only when the transmit power of the APs is at a moderate level (e.g. greater than $-5 \text{ dBm}$).
\subsubsection{Number of RIS Elements}
\begin{figure}[!t]
	\centering
	\includegraphics[width=0.7\textwidth]{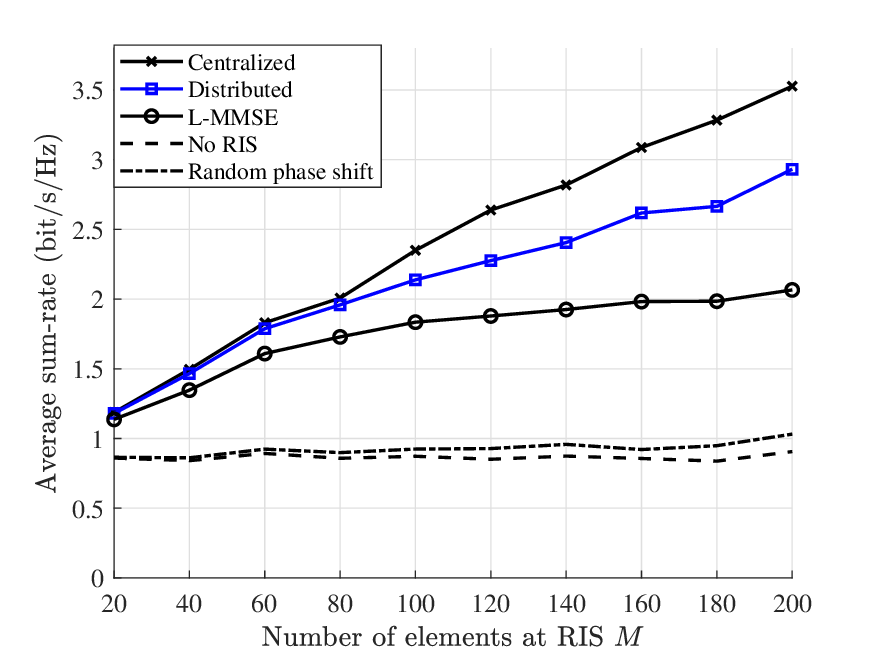}
	\caption{Average sum-rate versus the number of element at RIS $M$ with $I_{\rm o} = 20$.}
	\label{fig:WSRvsM}
\end{figure}

Adopting the same setups as above, the ASR versus the number of RIS elements is depicted in Fig. \ref{fig:WSRvsM}. We can observe that the ASR of the proposed joint distributed precoding and beamforming framework increases with the number of RIS elements. More importantly, we find that the performance gap between the centralized case and the distributed case is widen with the increasing number of RIS elements. This is because the size of feasible solution set increases with the number of RIS elements requiring more number of iterations for the proposed algorithm to converge. As such, for a fix number of $I_{\rm o}$, only a less effective solution to \eqref{equ:f_lk} can be obtained.
\subsubsection{Number of UE}
\begin{figure}[!t]
	\centering
	\includegraphics[width=0.7\textwidth]{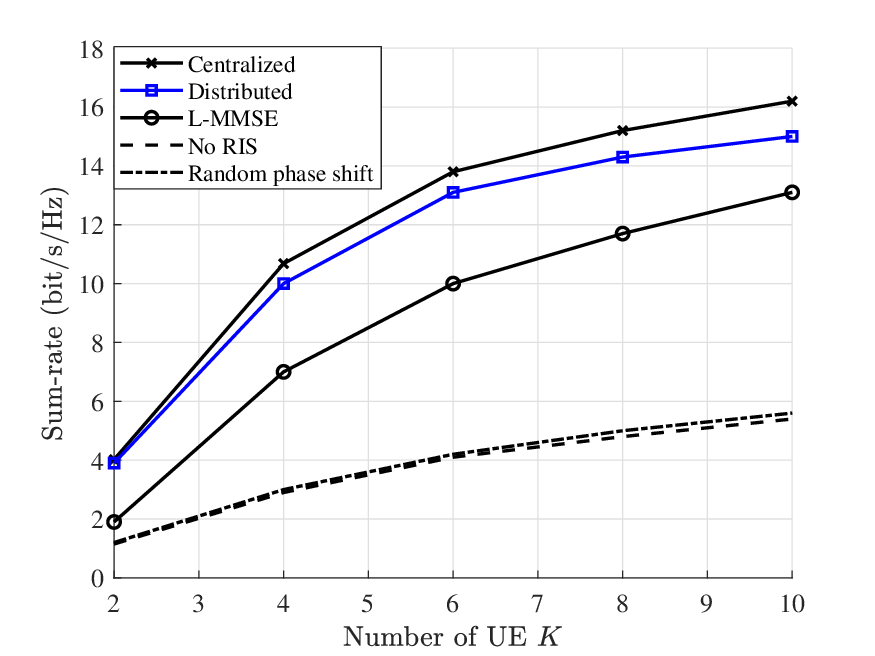}
	\caption{Sum-rate versus the number of UE $K$ with $I_{\rm o} = 20$.}
	\label{fig:SRvsK}
\end{figure}

Adopting the same configurations as described earlier, Figure \ref{fig:SRvsK} illustrates the sum-rate versus the number of UE elements. It is noticeable that as the number of UEs increases, the proposed distributed approach scales similarly to the centralized one. However, the extent of improvement diminishes with the increasing number of users. This is because the amplification of interference with the growthing number of users.

\section{Conclusion}
In this paper, we investigated a downlink RIS-aided cell-free network. We proposed a novel joint distributed precoding and beamforming framework to jointly design combining vectors, active precoding, and passive RIS beamforming. This framework decentralized the alternating optimization method to obtain a suboptimal solution with the goal of minimizing the weighted MSE. The algorithm complexity is reduced compared with the centralized algorithm. We demonstrated that the proposed distributed approach can achieve performance close to that of the centralized one, indicating the viability and efficiency of the proposed framework.


\bibliographystyle{IEEEtran}
\bibliography{./bibtex/BF_RIS_CF}


\end{document}